\documentclass[usenatbib]{mn2e}
\usepackage{psfig}

\newcommand{\etal}{et~al.}

\newcommand{\ionhy}{H{\sc ii}}

\newcommand{\transe}{$2_{1}\mbox{-}3_{0}\mbox{~E}$}

\newcommand{\degrees}{$^\circ$}
\newcommand{\kms}{$\mbox{km~s}^{-1}$}
\newcommand{\cc}{\mbox{$\,\rm cm^{-3}$}}
\newcommand{\ccs}{\mbox{$\,\rm cm^{-3} s$}}
\newcommand{\nH}{\mbox{$n_{\rm H}$}}
\newcommand{\scd}{\mbox{$N/\Delta V$}}
\newcommand{\scdm}{\mbox{$N_{\rm M}/\Delta V$}}
\newcommand{\scdh}{\mbox{$N_{\rm OH}/\Delta V$}}
\newcommand{\DV}{\mbox{$\Delta V$}}
\newcommand{\Td}{\mbox{$T_{\rm d}$}}
\newcommand{\Tk}{\mbox{$T_{\rm k}$}}
\newcommand{\Tb}{\mbox{$T_{\rm b}$}}
\newcommand{\Xm}{\mbox{$X_{\rm M}$}}
\newcommand{\Xh}{\mbox{$X_{\rm OH}$}}
\newcommand{\beam}{\mbox{$\epsilon^{-1}$}}
\newcommand{\HII}{H\,{\sc ii}}
\newcommand{\WHII}{\mbox{$W_{\rm HII}$}}
\newcommand{\Te}{\mbox{$T_{\rm e}$}}
\newcommand{\fe}{\mbox{$f_{\rm e}$}}

\newcommand{\lta}{\raisebox{-0.6ex}{$\,\stackrel
{\raisebox{-.2ex}{$\textstyle <$}}{\sim}\,$}}

\newcommand{\specdfig}[2]        
{
  \begin{center}
    \begin{minipage}[t]{0.45\textwidth}
        \psfig{file=#1.eps,height=0.75\textwidth,angle=270}
    \end{minipage}
    \hfill
    \begin{minipage}[t]{0.45\textwidth}
        \psfig{file=#2.eps,height=0.75\textwidth,angle=270}
    \end{minipage}
  \end{center}
}

\newcommand{\specsfig}[1]        
{
  \begin{center}
    \begin{minipage}[t]{0.45\textwidth}
        \psfig{file=#1.eps,height=0.75\textwidth,angle=270}
    \end{minipage}
  \end{center}
}
\newcommand{\speclfig}[2]        
{
  #1
  \begin{center}
    \begin{minipage}[t]{0.45\textwidth}
      \psfig{file=#2.eps,height=0.75\textwidth,angle=270}
    \end{minipage}
  \end{center}
}

\begin{document}


\title[19.9-GHz methanol maser emission] 
{Discovery of new 19.9-GHz methanol masers in star forming regions}

\author[Ellingsen \etal\/]{S.P. Ellingsen$^1$, D.M. Cragg$^2$, 
  J.E.J. Lovell$^3$, A.M. Sobolev$^4$, 
  \newauthor P.D. Ramsdale$^1$, P.D. Godfrey$^2$\\
$^1$ School of Mathematics and Physics, University of Tasmania, 
     Private Bag 21, Hobart, Tasmania 7001, Australia;\\  
     Simon.Ellingsen@utas.edu.au\\
$^2$ School of Chemistry, Building 23, Monash University, Victoria 3800, 
     Australia;\\
     Dinah.Cragg@sci.monash.edu.au, Peter.Godfrey@sci.monash.edu.au\\
$^3$ Australia Telescope National Facility, CSIRO, PO Box 76, Epping,
     NSW 2121, Australia;\\
     Jim.Lovell@csiro.au\\
$^4$ Astronomical Observatory, Ural State University, Lenin Street 51, 
     Ekaterinburg 620083, Russia;\\
     Andrej.Sobolev@usu.ru}

\maketitle

\begin{abstract}
  
  We have used the NASA Tidbinbilla 70-m antenna to search for emission
  from the \transe\/ (19.9-GHz) transition of methanol.  The search
  was targeted towards 22 star formation regions which exhibit maser
  emission in the 107.0-GHz $3_{1}\mbox{-}4_{0}\mbox{~A}^{+}$ methanol
  transition, as well as in the 6.6-GHz
  $5_{1}\mbox{-}6_{0}\mbox{~A}^{+}$ transition characteristic of
  class~II methanol maser sources.  A total of 7 sources were detected
  in the \transe\/ transition, 6 of these being new detections.  Many
  of the new detections are weak (\lta 0.5~Jy), however, they appear
  to be weak masers rather than thermal or quasi-thermal emission.
  
  We find a strong correlation between sources which exhibit 19.9-GHz
  methanol masers and those which both have the class~II methanol masers
  projected against radio continuum emission and have associated
  6035-MHz OH masers.  This suggests that the 19.9-GHz methanol masers
  arise in very specific physical conditions, probably associated with
  a particular evolutionary phase.  In the model of \citet*{CSG02}
  these observations are consistent with gas temperatures of 50~K,
  dust temperatures of 150-200~K and gas densities of
  $10^{6.5}-10^{7.5}$ \cc.

\end{abstract}


\section{Introduction}

In the 20 years since the discovery of the first of the class II methanol
maser transitions \citep{WWSJ84} they have become important probes of
high-mass star formation regions.  Class~II methanol masers have a
number of advantages over the other molecular maser species commonly
found in star forming regions.  They are more common than main-line OH
masers and where they are found towards the same regions they are
typically an order of magnitude stronger \citep{CVEWN95}.  They are
significantly less variable than 22-GHz water masers, with individual
features having lifetimes in excess of a decade and unlike OH and
water masers they appear to be exclusively associated with high-mass
star formation \citep{MENB03}.  

To date more than 20 different methanol transitions have been observed
as interstellar masers, and these have been empirically divided into
two classes \citep{M91a}.  Class~II methanol masers are closely
associated with high-mass star formation and other tracers of this
process, far-infrared sources, \ionhy\ regions and OH and water
masers.  Class~I masers are also found within high-mass star forming
regions, but offset from the young stellar objects and are thought to
be collisionally pumped in outflows.  The most common class~II
methanol maser transitions are those at 6.6 and 12.1~GHz
\citep{M91b,BMMW87}, but some sources such as W3(OH), NGC6334F and
345.010+1.792 exhibit methanol maser emission from many other
transitions \citep{CSECGSD01,SSECMOG01}.  Interferometric observations
have shown that the 6.6- and 12.1-GHz emission in some sources is
coincident both in velocity and spatially (to within a few
milliarcseconds) \citep*{MRPMW92,NWCWG93,MBC00}.  Observations with
arcsecond resolution of some of the rarer transitions in W3(OH) show
that to within the relative positional accuracy these are also
spatially coincident with the 6.6-GHz masers
\citep{MJWWW88,SSECMOG01}.

The pumping of interstellar masers is a complicated process that
sensitively depends upon a number of factors such as the gas
temperature and density, molecular abundance and external radiation
field.  Theoretical studies of maser pumping show that the commonly
observed maser transitions are produced over a wide range of physical
conditions \citep{CSG02,GFD92,PK96a,PK96b,SCG97} and hence it isn't
possible to infer any detailed information about the physical
conditions in the maser region from observations of a single
transition.  However, where multiple transitions arise from the same
gas any maser pumping scheme must be able to simultaneously produce
the various transitions in the observed intensity ratio.  This
approach has been applied to OH and methanol maser emission in a small
number of sources and produces physically plausible answers
\citep{CW91,CSECGSD01,SSECMOG01}.  Typically the star-forming regions
targetted in searches for new maser transitions and multi-transition
modelling are those which exhibit emission from a large number of
maser transitions such as W3(OH) and NGC6334F.  However, the presence
of a large number of different maser transitions in these sources
indicates that these regions are special in some way, and not
representative of the conditions that exist in the majority of
high-mass star forming regions.  In order to obtain information on the
conditions in a larger sample of more representative star forming
regions we are undertaking a program to search for methanol maser
transitions that are rarely strong, (hence forth referred to as ``weak
methanol'' maser transitions for simplicity) towards all sources that
exhibit 107.0-GHz methanol maser emission.  To date 25 such sources
have been identified from searches of more than 175 class~II methanol
maser sources \citep{VDKSBW95,VESKOV99,CYBC00,MB02}.  In constraining
the physical conditions in these regions the non-detection of the weak
methanol maser transitions can sometimes be as useful as a detection,
as it also limits the range of parameter space within the model
consistent with observations of the region \citep[see for
example][]{CSCEG04}.

We have previously undertaken searches for 23.1-, 85.5-, 86.6- and
108.8-GHz methanol masers towards the sample of star formation regions
with 107.0-GHz methanol masers \citep{CSCEG04,ECMMG03,VESKOV99}.
Although it was one of the first methanol maser transitions
discovered, few observations have been made of the 19.9-GHz \transe\/
transition and it has only been reported towards three star forming
regions, W3(OH), NGC7538 \citep{WWMH85} and NGC6334F \citep{MB89}.  Here
we report the results of a sensitive search for maser emission from
the \transe\/ transition of methanol with the Tidbinbilla 70-m
antenna.  The search was targetted towards the 22 known 107.0-GHz
methanol maser sources that are observable from Tidbinbilla.

\section{Observations and Data reduction}

The observations of the 19.9-GHz \transe\/ transition were made on
2003 April 10, October 25 \& 26 and November 15-17 using the NASA
Tidbinbilla 70-m antenna.  The rest frequency assumed was 19967.3961
MHz, which has a 2$\sigma$ uncertainty of 0.0002~MHz \citep*{MDM85}.  At
19.9~GHz the FWHM antenna beamwidth was 53\arcsec\/.  The sources were
observed in position switching mode with 300 seconds spent at the
onsource position and 150 seconds spent at a position offset by +2.5
minutes in right ascension prior to the onsource observation and
another 150 seconds spent at a position offset by -2.5 minutes in
right ascension following the observation.  This observing pattern
means that the movement of the antenna during the observation is the
same for both the onsource and reference observations and minimises
the difference in atmospheric and antenna gain effects between them.
The weather conditions varied considerably during the observations and
the majority of the sources were observed on more than one occasion.
The observations made in good weather conditions typically have a
sensitivity a factor of two or more better than those observed in poor
conditions.  For sources observed in good weather conditions an
onsource integration time of 5 minutes typically yielded an RMS of
0.1~Jy.  The data for a single circular polarization (LCP) were
collected using a 2-bit, three level digital autocorrelation
spectrometer configured with 4096 channels spanning a 16-MHz
bandwidth.  For an observing frequency of 19.9 GHz this configuration
produces a natural weighting velocity resolution of 0.07~\kms, or
0.12~\kms\/ after Hanning smoothing.  The total velocity range covered
was approximately 240~\kms, and for each source this centred around
the velocity of the 6.6-GHz methanol maser emission.  Tipping curves
to measure the opacity of the atmosphere were performed for the
observations made in November.  For the earlier observations the
opacity was estimated either from the variation in the system
temperature with elevation, or from tipping scans made at other
frequencies.  The nominal pointing accuracy of the Tidbinbilla 70-m
antenna is 10\arcsec\/ and comparison of sources observed on more than
one occasion suggests that the final absolute flux density
calibration is accurate to 10\%

For some of the sources detected at 19.9 GHz, additional observations
at 6.6 and 12.1~GHz were made with the University of Tasmania 26-m
antenna at Hobart on 2003 December 1 \& 2.  At 6.6- and 12.1-GHz the
FWHM beamwidths were 7.1\arcmin\/ and 3.9\arcmin\/ respectively.  At
both frequencies the observations were made with a dual circularly
polarized receiver with a typical system equivalent flux density of
880 Jy at 6.6~GHz and 790~Jy at 12.1~GHz.  The data were collected
using a 2-bit, three level digital autocorrelation spectrometer
configured with 4096 channels spanning a 4-MHz bandwidth at 6.6~GHz
and 2048 channels spanning an 8-MHz bandwidth at 12.1~GHz.  This
configuration produces a natural weighting velocity resolution of
0.05~\kms, or 0.09~\kms\ after Hanning smoothing at 6.6~GHz, and
natural weighting and Hanning smoothed velocity resolutions of 0.12
and 0.19~\kms\/ at 12.1~GHz.  This velocity resolution is well matched
to that of the 19.9-GHz observations.  The sources were observed in
position switching mode with 600 seconds spent at a position offset in
position by -1 degree in declination and 600 seconds spent onsource at
6.6~GHz and 60 seconds spent onsource at 12.1~GHz.  In addition a five
point grid with observations at the nominal position and at four
points each offset by approximately half the FWHM beamwidth were made
for each source at both frequencies.  These data were used to measure
any pointing offset and the flux density of each source has been
scaled to correct for the offset.  Analysis of the grid observations
revealed that the 12.1~GHz observations were made with the receiver
significantly offset from the optimal position.  The absolute flux
density calibration of the 6.6-GHz observations is better than 10\%,
however, uncertainties due to the offset receiver position mean that
at 12.1~GHz the calibration is probably only good to a factor of 2.

\section{Results}

A total of 22 sources were observed at 19.9~GHz and these are listed
in Table~\ref{tab:meth19}.  Emission from the 19.9-GHz \transe\/
transition was detected towards 7 sources, all of these except
NGC6334F are new discoveries.  The 19.9-GHz spectrum of each of the
detected sources is shown in Fig.~\ref{fig:meth19}.  For the sources
where emission was detected Gaussian profiles were fitted and the
parameters of these are listed in Table~\ref{tab:meth19}, for the
other sources a limit of three times the RMS noise in the final
spectra is given.  The typical 3-$\sigma$ limit for these observations
was less than 0.3~Jy which is significantly better than our searches
for the 23.1-, 85.5-, 86.6- and 108.8-GHz transitions.  Of the 7
sources detected only two have a peak flux density stronger than 1~Jy
and so these observations demonstrate the importance of sensitivity in
searches for weak methanol maser transitions.  The weakness of the 19.9-GHz
methanol emission in most of our detections means that we cannot
conclusively prove that it is maser emisison, however, comparison with
other methanol maser transitions in each source shown in
Figs~\ref{fig:stack323}-\ref{fig:stack353} strongly suggests that for
the majority it is.  The recent upgrade to the Australia Telescope
Compact Array (ATCA) means that it is now able to observe the 19.9-GHz
methanol transition and hence will be able to determine whether the
weak sources are masers.

\begin{table*}
  \caption{A list of sources observed at 19.9-GHz.  For sources where 
  emission was detected the values listed are those of Gaussian profiles
  fitted to the spectra.  For sources where no emission was detected the 
  value listed is 3 x the RMS noise level in the Hanning smoothed spectrum.
  DOY 100 = 10 April 2003 ; DOY 299 = 26 October 2003 ; DOY 319 = 15 November
  2003}  
  \begin{tabular}{lllrrccc} \hline
    {\bf Name} & {\bf R.A.(J2000)} & {\bf Dec.(J2000)}                      & 
    {\bf Observed} & {\bf Integration} & {\bf Flux Density} & {\bf Velocity} &
    {\bf FWHM}     \\
               & {\bf (h~m~s)}     & {\bf (\degrees\/ \arcmin\/ \arcsec\/)} & 
    {\bf DOY 2003} & {\bf Time (min)}  & {\bf (Jy)}         & {\bf (\kms\/)} &
    {\bf (\kms\/)} \\ [2mm] \hline
    188.946+0.886  & 06:08:53.3 & +21:38:29  & 100         & 10 & $<$0.20 &
           & \\
    192.600--0.048 & 06:12:54.0 & +17:59:24  & 100         & 10 & $<$0.19 &
           & \\
    310.144+0.760  & 13:51:58.5 & --61:15:42 & 300,320     & 10 & $<$0.16 &
           & \\
    318.948--0.196 & 15:00:55.4 & --58:58:53 & 300,320     & 10 & $<$0.17 &
           & \\
    323.740--0.263 & 15:31:45.5 & --56:30:50 & 299,319     & 10 & 0.15    &
     -51.4 & 3.1 \\
    327.120+0.511  & 15:47:32.7 & --53:52:38 & 300,320     & 10 & $<$0.17 &
           & \\
    328.808+0.633  & 15:55:48.5 & --52:43:07 & 300,319,320 & 15 & 0.79    &
     -43.7 & 0.9 \\
    336.018--0.827 & 16:35:09.3 & --48:46:47 & 320         &  5 & $<$0.21 &
           & \\
    339.884--1.259 & 16:52:04.7 & --46:08:34 & 299,319,320 & 15 & 9.8     & 
     -36.2 & 0.4 \\
                   &            &            &             &    & 1.5     & 
     -36.8 & 0.4 \\
    340.054--0.244 & 16:48:13.9 & --45:21:44 & 320         &  5 & $<$0.20 &
           &     \\
    340.785--0.096 & 16:50:14.8 & --44:42:26 & 320         &  5 & $<$0.76 &
           &     \\
    345.003--0.223 & 17:05:10.9 & --41:29:06 & 321         &  5 & 0.22    &
     -27.4 & 3.4 \\
    345.010+1.792  & 16:56:47.6 & --40:14:26 & 320         &  5 & 0.21    & 
     -17.6 & 1.5 \\
    345.504+0.348  & 17:04:22.9 & --40:44:22 & 300         &  5 & $<$0.65 &
           &     \\
    348.703--1.043 & 17:20:04.1 & --38:58:31 & 320         &  5 & $<$0.39 &
           &     \\
    NGC6334F       & 17:20:53.4 & --35:47:01 & 320         &  5 & 146.7   & 
     -10.4 & 0.3 \\
                   &            &            &             &    & 26.8    & 
     -11.0 & 0.3 \\
                   &            &            &             &    & 1.9     & 
     -9.6  & 0.4 \\
    353.410--0.360 & 17:30:26.2 & --34:41:46 & 320,321     &  7 & 0.33    & 
     -20.5 & 1.8 \\
    9.621+0.196    & 18:06:14.7 & --20:31:32 & 320,321     & 12 & $<$0.22 &
           &     \\
    12.909--0.260  & 18:14:39.5 & --17:52:00 & 320         & 10 & $<$0.27 &
           &     \\
    23.010--0.411  & 18:34:40.3 & --09:00:38 & 320         & 10 & $<$0.26 &
           &     \\
    23.440--0.182  & 18:34:39.2 & --08:31:24 & 320         & 10 & $<$0.24 &
           &     \\
    35.201--1.736  & 19:01:45.5 & +01:13:35  & 320         & 10 & $<$0.51 &
           &     \\ \hline
  \end{tabular}
  \label{tab:meth19}
\end{table*}

\begin{figure*}
  \specdfig{efig1a}{efig1b}
  \specdfig{efig1c}{efig1d}
  \specdfig{efig1e}{efig1f}
  \specsfig{efig1g}
\caption{Spectra of the sources detected at 19.9 GHz, all spectra have 
  been Hanning smoothed, except NGC6334F.}
\label{fig:meth19}
\end{figure*}

\subsection{Comments on individual sources} \label{sec:indiv}

\subsubsection*{323.740-0.263}

\begin{figure}
  \psfig{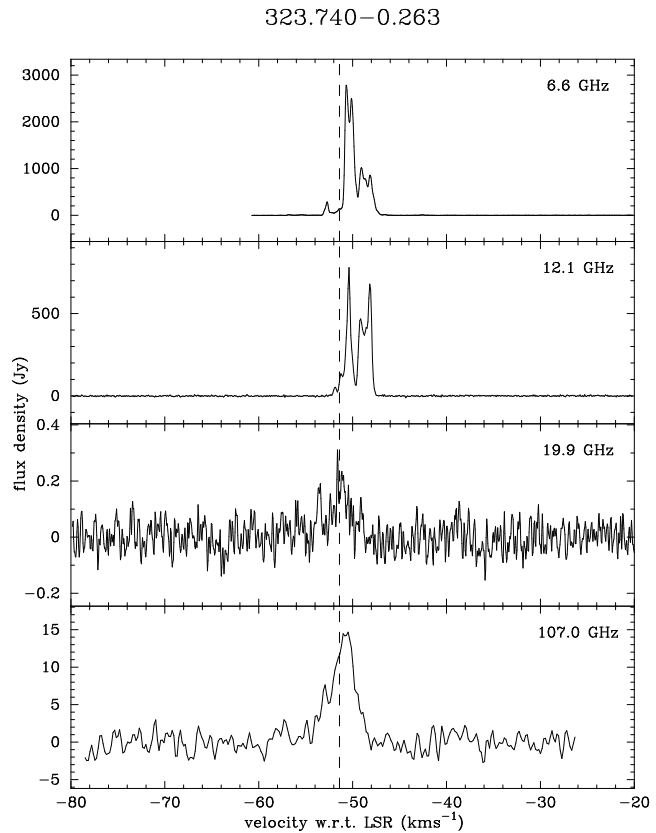}
  \caption{6.6-, 12.2-, 19.9- and 107.0-GHz methanol emission towards 
    323.740-0.263.  The 107.0-GHz spectrum is from \protect\citet{VESKOV99}.}
\label{fig:stack323}
\end{figure}

This is one of the strongest 6.6- and 12.1-GHz methanol masers,
however at 19.9~GHz it was only marginally detected with a peak flux
density of 0.15 Jy and a FWHM of 3.1~\kms.  The velocity and width of
the 19.9-GHz emission is similar to that seen at both 107.0- and
156.6-GHz \citep{CYBC00}.  Caswell \etal\/ interpret the strong
107.0-GHz emission as being due to a blend of maser lines, with some
thermal contribution and the 156.6-GHz emission as being thermal.
Fig.~\ref{fig:stack323} shows that the 19.9-GHz emission appears to be
slightly offset from the strongest 6.6- and 12.1-GHz masers, although
the low signal-to-noise ratio means that it is not possible to be
definitive about this.  The weak, broad nature of the 19.9-GHz
emission in this source and its alignment with thermal emission from
other methanol transitions \citep{CYBC00} means that it may too be
thermal, but higher spatial resolution observations are required to
determine this.

323.740-0.263 is an unusual star forming region in that although it
shows extremely strong class~II methanol maser emission \citet{PNEM98}
did not detect centimetre radio continuum emission with a 4-$\sigma$
limit of 0.2 mJy/beam.  Infra-red emission is present towards
323.740-0.263, but only at wavelengths in excess of 3.6~\micron\/
\citep*{DPT00,WLB02}.  From their 10 and 18~\micron\/ observations De
Buizer \etal\ calculate a dust temperature of 128~K.  Walsh \etal\/
also detected H$_2$ emission from the region surrounding the masers,
some of which appears to be shocked excited.

\subsubsection*{328.808+0.633} 

\begin{figure}
  \psfig{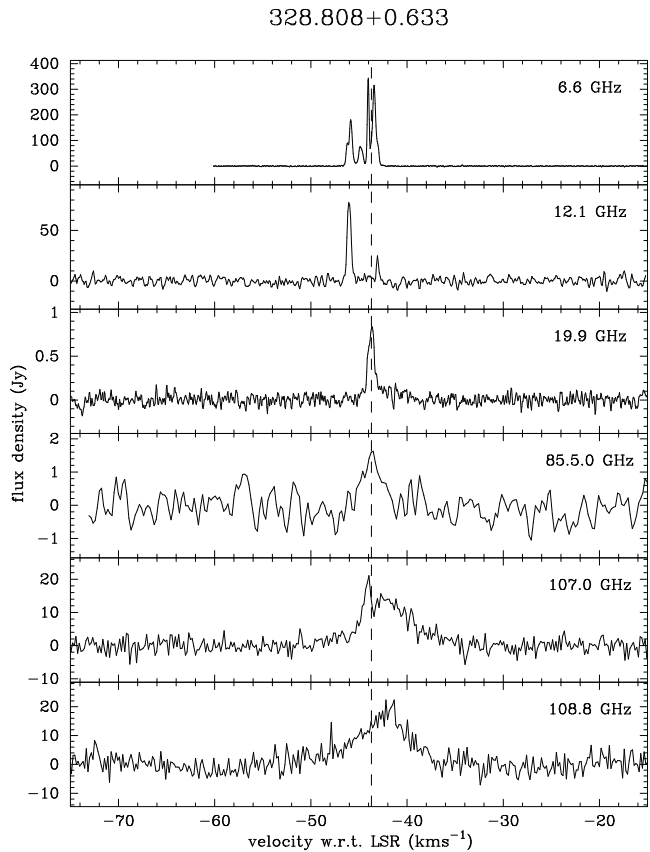}
  \caption{6.6-, 12.2-, 19.9-, 85.5-, 107.0- and 108.8-GHz methanol emission
    towards 328.808+0.633.  The 85.5-GHz spectrum is from
    \protect\citet{ECMMG03} and the 107.0- and 108.8-GHz spectra are
    from \protect\citet{VESKOV99}.}
\label{fig:stack328}
\end{figure}

Fig.~\ref{fig:stack328} shows that the 19.9-GHz methanol maser
emission in this source has the same peak velocity as the weak
85.5-GHz emission detected by \citet{ECMMG03}, which is slightly
offset in velocity from the strongest 6.6- and 107.0-GHz emission.
The 12.2-GHz emission at velocities near -44~\kms\/ is not detected in
our Mt Pleasant observations, but \citet{CVEN95} found emission with a
strength of 1-2 Jy in this velocity range (beneath the detection
threshold of the Mt Pleasant observations).  The emission at 107.0-GHz
has both maser and thermal components, while the emission at 108.8-
and 156.6-GHz is thought to be thermal \citep{CYBC00,VESKOV99}, and is
offset from the velocity of the 19.9-GHz methanol maser emission.

The class~II methanol masers in 328.808+0.633 are projected against a
strong cometary \ionhy\/ region \citep{WBHR98} and 1720-, 4765-, 6030-
and 6035-MHz OH masers have all been detected within a few arcseconds
of the methanol masers \citep{C03,C04,DE02}.  The masers are also
associated with a mid-infrared source, from which a dust temperature
of approximately 120~K is inferred \citep{DPT00}.

\subsubsection*{339.884-1.259}

\begin{figure}
  \psfig{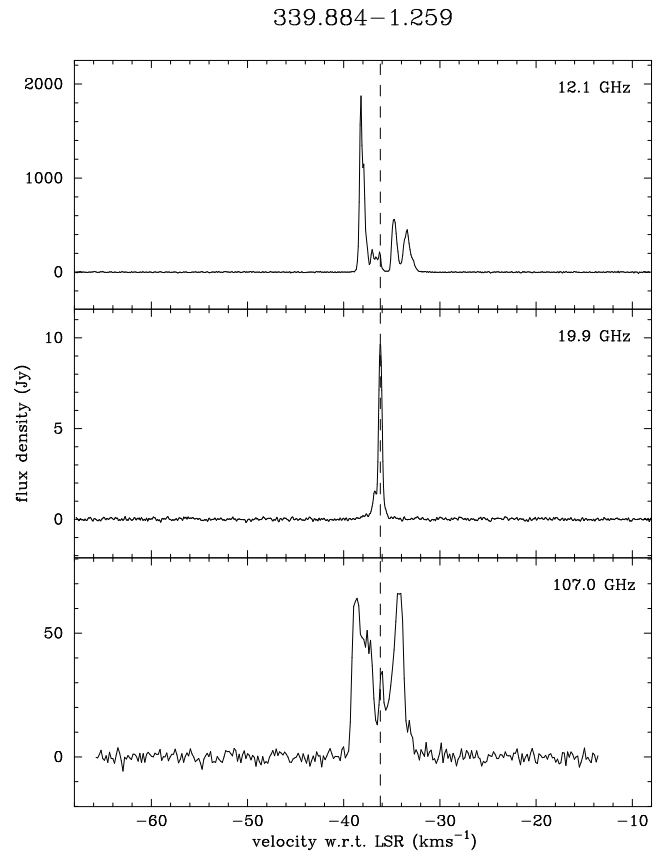}
  \caption{12.2-, 19.9- and 107.0-GHz methanol emission towards 339.884-1.259.
    The 107.0-GHz spectrum is from \protect\citet{VESKOV99}.}
\label{fig:stack339}
\end{figure}

The strong, narrow emission at 19.9~GHz in this source identifies it
unambiguously as maser emission.  The 19.9-GHz methanol masers are at
velocities in the middle of the range seen in the 6.6- and 12.1-GHz
masers.  Fig.~\ref{fig:stack339} shows that the strongest 19.9-GHz
methanol maser emission in 339.884-1.259 is at a velocity
corresponding to one of the weaker peaks in the 12.1- and 107.0-GHz
emission.  The observations of \citet{CVEWN95} show that at the
corresponding velocity the 6.6-GHz is at a minimum.  Very long
baseline interferometry observations by \citet{P98} show a narrow
6.6-GHz emission feature with a flux density of approximately 300~Jy
at a velocity of -36.1~\kms.  The observations of \citet{CYBC00}
suggest that both the 107.0 and weaker 156.6-GHz methanol masers are
superimposed on top of thermal emission and show significant
variations on a timescale of years.

The class~II methanol masers in this source are projected against the
centre of a weak cometary \ionhy\/ region \citep*{ENM96} and 1720-,
6030- and 6035-MHz OH masers have all been detected within a few
arcseconds of the methanol masers \citep{C03,C04}.  \citet{DWPPT02}
detect three mid infra-red sources close to the masers with a peak
dust temperature of 145~K and suggest that some of the radio continuum
emission results from a collimated ionized outflow.

\subsubsection*{345.003-0.223}

\begin{figure}
  \psfig{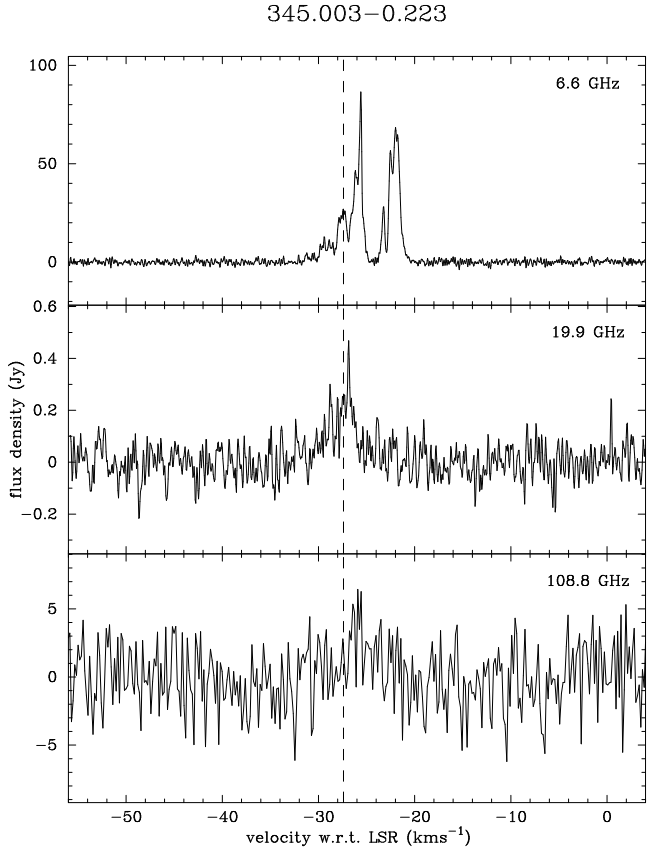}
  \caption{6.6-, 19.9- and 108.8-GHz methanol emission towards 345.003-0.223.
    The 108.8-GHz spectrum is from \protect\citet{VESKOV99}.}
\label{fig:stack3450}
\end{figure}

The 19.9-GHz methanol masers in 345.003-0.223 follow the pattern of
the majority of the other sources in being offset in velocity from the
strongest emission at 6.6-GHz (see Fig.~\ref{fig:stack3450}), but
coincident in velocity with a weaker emission feature.  However, the
peak flux densities of the 6.6-, 12.1- and 107.0-GHz masers in this
source are significantly less than for the other sources
\citep{CVEN95,CYBC00}.  The velocity of the 19.9-GHz peak lies between
the velocity of the thermal emission at 156.6-GHz (-27.8~\kms) and the
peak of the 107.0-GHz maser emission (-26.9~\kms), which also lies on
top of broad thermal emission \citep{CYBC00}.  The 12.1-GHz maser
emission from the velocity range of the 19.9-GHz maser is very weak
(of the order of 1~Jy) \citep{CVEN95} and comparison of the 6.6-GHz
emission shown by \citet{CVEWN95} with that in
Fig.~\ref{fig:stack3450} shows that the 6.6-GHz peak flux density is
less than 20\% of that observed a decade ago.

\citet{WBHR98} found the 6.6-GHz methanol masers in this source to lie
in two clusters, the emission near -27~\kms\/ is projected against an
\ionhy\/ region and 1720-, 6030- and 6035-MHz OH masers have all been
detected within a few arcseconds of the class~II methanol masers
\citep{C03,C04}

\subsubsection*{345.010+1.792} 

\begin{figure}
  \psfig{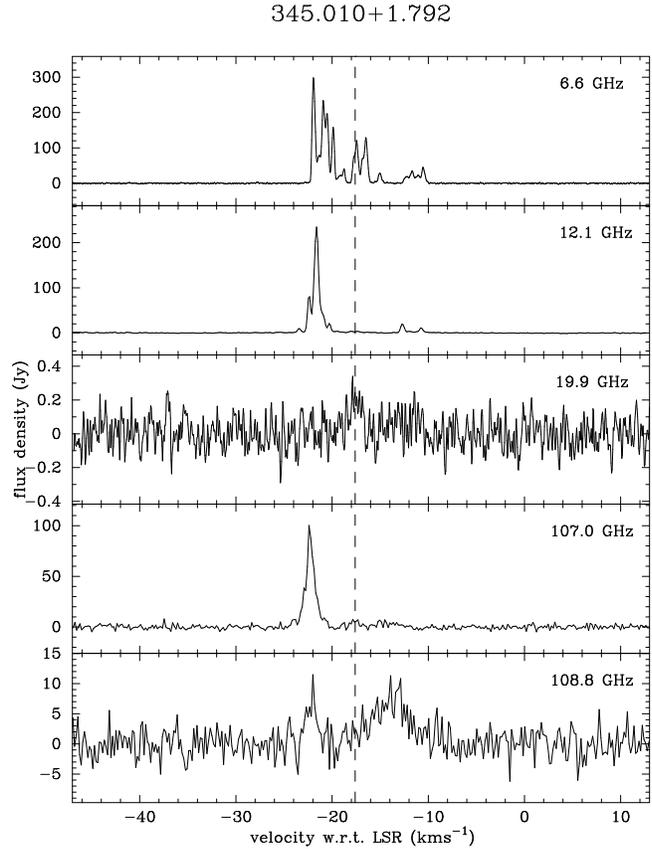}
  \caption{6.6-, 12.2-, 19.9-, 107.0- and 108.8-GHz methanol emission
    towards 345.010+1.792.  The 12.2-GHz spectrum is from
    \protect\citet{CSECGSD01} and the 107.0- and 108.8-GHz spectra are
    from \protect\citet{VESKOV99}.}
\label{fig:stack345}
\end{figure}

345.010+1.792 is a remarkable source which exhibits class~II methanol
maser emission in nearly every known transition in which it has been
observed \citep{V98,VESKOV99,CSECGSD01}, to date the only exception is
the 23.1-GHz transition \citep{CSCEG04}.  However, as can be seen in
Fig.~\ref{fig:stack345} the 19.9-GHz methanol maser emission differs
from the other weak methanol transitions which are generally observed at
velocities near -22~\kms\ \citep{CSECGSD01}.  The 19.9-GHz maser
emission at -17.6~\kms\ corresponds to a 100~Jy 6.6-GHz emission
feature, which at one time was the strongest feature in the spectrum,
but has significantly declined in strength over the last decade
\citep{CVEWN95}.  There is 12.1- and 107.0-GHz emission present at the
velocity of the 19.9-GHz emission, but only at a level of a few Jy
\citep{CYBC00}.  Strong thermal emission (most evident in the 108.8-GHz
spectrum) is also present in many of the millimetre methanol transitions that
have been observed towards 345.010+1.792, however, it is well offset
in velocity from the 19.9-GHz peak and the other weak methanol maser
transitions.

The class~II methanol masers are projected against an \ionhy\/ region,
although significantly offset from the centre \citep{WBHR98}.  The 10
and 18~\micron\/ MIR emission associated with the masers implies a dust
temperature of approximately 160~K \citep{DPT00} and 6030- and 6035-MHz
OH masers are located within a few arcseconds of the methanol
\citep{C03}.

\subsubsection*{NGC6334F} 

\begin{figure}
  \psfig{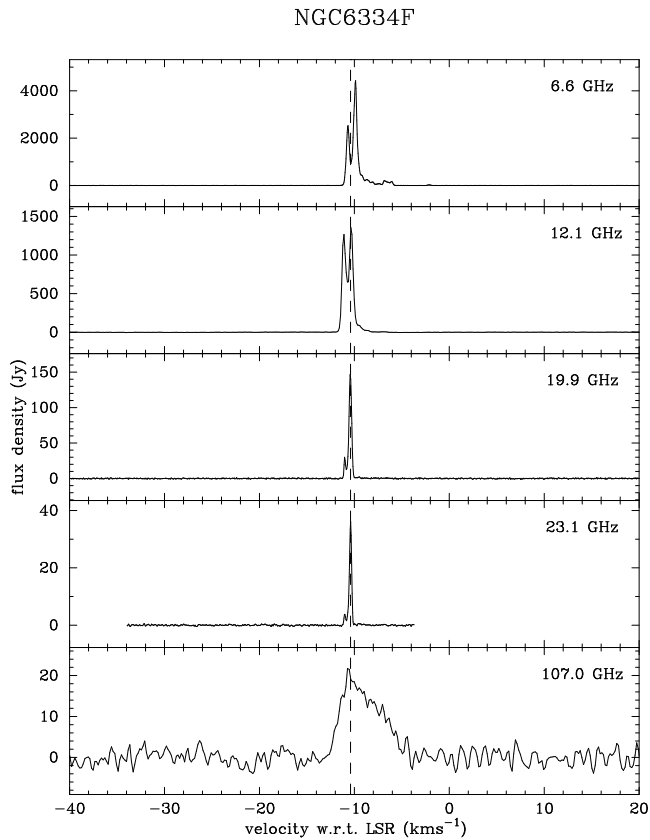}
  \caption{6.6-, 12.2-, 19.9-, 23.1- and 107.0-GHz methanol emission
  towards NGC6334F.  The 12.2-GHz spectrum is from \protect\citet{CSECGSD01},
  the 23.1-GHz spectrum is from \protect\citet{CSCEG04} and the 107.0-GHz 
  spectrum is from \protect\citet{VESKOV99}.}
\label{fig:stackngc}
\end{figure}

NGC6334F is the archetypal high-mass star forming region of the
southern sky, it shares many characteristics with W3(OH) and has been
well studied over a range of wavelengths and molecular transitions.
It remains the strongest known 19.9-GHz methanol maser source,
although it has significantly decreased in strength over the 16 year
period since its discovery \citep{MB89}.  Unlike most of the other
19.9-GHz methanol maser sources Fig.~\ref{fig:stackngc} shows that in
NGC6334F the 19.9-GHz peak is at the same velocity as the peak flux
density at 6.6-, 12.1- and 107.0-GHz.  There are two regions of
class~II methanol maser emission near NGC6334F, separated by about 6
arcseconds, one projected onto the \ionhy\/ region, the other offset
from it \citep{ENM96}.  Unfortunately both clusters show strong
emission at velocities around -10.4~\kms\/ and interferometric
resolution observations are required to determine which of the
clusters the 19.9-GHz emission is associated with.

NGC6334F has 1720-, 6030- and 6035-MHz OH masers within a few
arcseconds of the class~II methanol maser cluster projected onto a
strong \ionhy\/ region \citep{C03,C04}.  There are a number of MIR
sources in the NGC6334F region, the strongest of which coincides with
the peak of the radio continuum and maser cluster \citep{DPT00}.  De
Buizer \etal\/ estimate a dust temperature of 159~K from their
observations.

\subsubsection*{353.410-0.360}

\begin{figure}
  \psfig{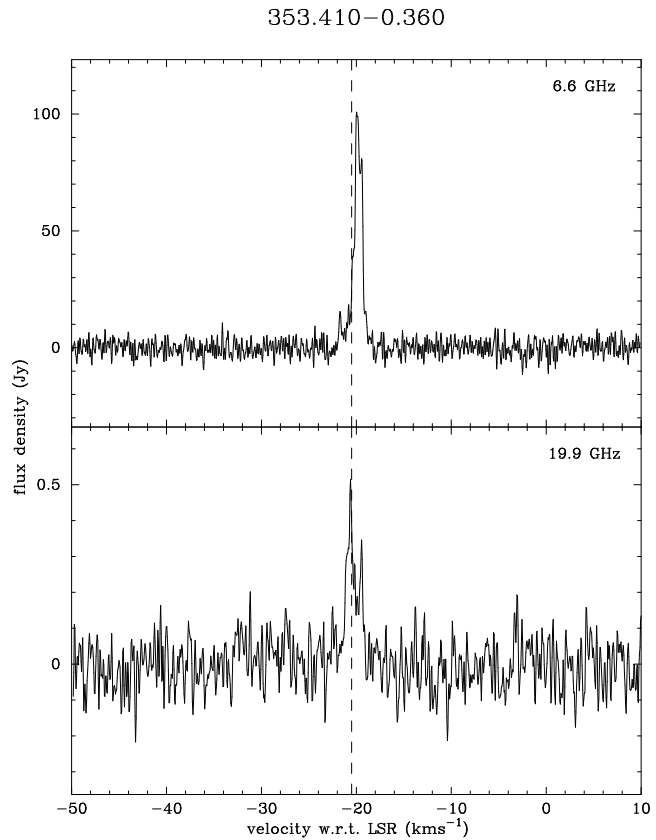}
  \caption{6.6- and 19.9-GHz methanol emission towards 353.410-0.360}
\label{fig:stack353}
\end{figure}

Fig.~\ref{fig:stack353} shows that the 19.9-GHz methanol maser
emission in 353.410-0.360 follows the pattern seen in most of the
other sources, being aligned in velocity with secondary 6.6-GHz and
12.1-GHz peaks \citep{CVEN95}.  The class~II methanol masers are
projected onto an \ionhy\/ region with a peak flux density of 215
mJy/beam at 3-cm \citep{PNEM98} and 1720-, 4765-, 6030- and 6035-MHz
OH masers are located within a few arcseconds \citep{C03,C04,DE02}.

\section{Discussion}

Table~\ref{tab:fluxes} presents a nearly complete set of observations
of 25 class~II methanol maser sources in 9 transitions.  The selected
sources all exhibit masers at 6.6 and 107.0~GHz, and all but 1 also
have masers detected at 12.1~GHz (although in some cases not at the
velocity of the 107.0~GHz peak).  Of the remaining 6 frequencies
surveyed, the 19.9-GHz maser which is the subject of this paper is
detected in the largest number of sources (9), followed by the
156.6-GHz maser (4~sources), the 23.1-GHz maser (3~sources), the 85.5-
and 86.6-GHz masers (2~sources each), and the 108.8-GHz maser
(1~source).  While these statistics demonstrate that these other maser
transitions are rarely strong, they suffer from a number of
limitations.  The searches at different frequencies were carried out
with different sensitivities, some are incomplete, and the
detectability of weak millimetre-wavelength masers is often
compromised by overlapping thermal emission.  Nevertheless, the
general scarcity of the weaker masers helps delimit the conditions
prevalent in our sample of 107.0-GHz maser sources, as is outlined in
section~\ref{sec:modelling}.

\begin{table*}
  \vbox to220mm{\vfil Landscape table to go here
  \caption{}
  \vfil}
  \label{tab:fluxes}
\end{table*}

\subsection{Association with radio continuum and 6-GHz OH masers}

It is clear from the comments on the individual sources that most
19.9-GHz methanol masers are :
\begin{itemize}
\item offset in velocity from the strongest 6.6- and 12.1-GHz masers ;
\item occur in sources that are projected against centimetre continuum 
  emission ;
\item have associated 6030- and 6035-MHz OH masers.
\end{itemize}
Our observations include 7 of the 9 sources known to exhibit 19.9-GHz
methanol maser emission (the other two being W3(OH) and NGC7538).
Both W3(OH) and NGC7538 also have the masers projected against
centimetre radio continuum \citep{ARM00,KM98} and associated 6035-MHz
OH masers \cite{BDWC97}.  We have examined the literature to look at
the association with radio continuum and OH maser transitions for all
25 sources that show 107.0-GHz methanol masers, the results are
summarised in Table~\ref{tab:assoc}.  This shows that there is a very
good correlation between centimetre radio continuum and 6035-MHz OH
masers and 19.9-GHz methanol masers.  Most of the 19.9-GHz detections
(7 of 9) are also associated with 1720-MHz OH maser emission, which is
only detected in about 7\% of OH maser sources in star forming regions
\citep{C98}.  There is only one source (323.740-0.263) which lacks
both radio continuum emission and 6035-MHz OH masers and has 19.9-GHz
methanol and the weak, broad nature of the emission in this source
means that it may be thermal.  Further, there is also only one source
(35.201-1.736/W48) which has the class~II methanol masers projected
against radio continuum emission, has associated 6035-MHz OH masers
but does not show 19.9-GHz methanol emission.  However, the absolute
position of the class II methanol masers with respect to the radio
continuum emission has not been accurately determined in this source
and so it may not be a true exception.  As all the sources in our
sample are 107.0-GHz methanol masers further observations are required
to determine if 19.9-GHz methanol masers are found more generally
towards sources projected against centimetre radio continuum with
6035-MHz OH masers.  We can use the observed association between
19.9-GHz methanol, 6035-MHz OH and radio continuum and the
\citet{CSG02} model of methanol and OH masers to constrain the
conditions in these sources.

\begin{table*}
  \caption{The association of 107.0-GHz methanol masers with centimetre
  radio continuum emission (are the class~II masers projected against the
  radio continuum) and OH masers.  References :
  1=\protect\citet{ARM00};
  2=\protect\citet{BDWC97};
  3=\protect\citet{C98};
  4=\protect\citet{C03};
  5=\protect\citet{C04};
  6=\protect\citet{CH83};
  7=\protect\citet{DRGF82};
  8=\protect\citet{DE02};
  9=\protect\citet{ENM96};
  10=\protect\citet{FC00};
  11=\protect\citet{GM87};
  12=\protect\citet{KM98};
  13=\protect\citet{MBC00};
  14=\protect\citet{OPB94};
  15=\protect\citet{PNEM98};
  16=\protect\citet*{SKH00};
  17=\protect\citet{WBHR98}}
  \begin{tabular}{lccrr} \hline
    {\bf Name} & \multicolumn{3}{c}{{\bf Associated with}} & 
      {\bf References} \\
               & {\bf cm continuum} & {\bf 19.9 GHz methanol}  
               & {\bf OH maser transitions} &                   \\ \hline
    W3(OH)         & Y & Y & 1612,1665,1667,1720,4765,6030,6035 & 2,11,12,16 \\ 
    188.946+0.886  & N & N & 1665                          & 1 \\
    192.600--0.048 & N & N &                               & 13 \\
    310.144+0.760  &   & N & 1665,1667                     & 3 \\
    318.948--0.196 & N & N & 1665,1667                     & 3,9 \\
    323.740--0.263 & N & Y & 1665,1667                     & 3,15 \\
    327.120+0.511  & Y & N & 1665,1667                     & 3,15 \\
    328.808+0.633  & Y & Y & 1665,1667,1720,4765,6030,6035 & 3,4,5,8,17 \\
    336.018--0.827 &   & N & 1665,1667                     & 3 \\
    339.884--1.259 & Y & Y & 1665,1667,1720,6030,6035      & 3,4,5,9 \\
    340.054--0.244 & N & N & 1665,1667                     & 3,17 \\
    340.785--0.096 & N & N & 1665,1667,1720,6035           & 3,4,5,17 \\
    345.003--0.223 & Y & Y & 1665,1667,1720,6030,6035      & 3,4,5,17 \\
    345.010+1.792  & Y & Y & 1665,1667,6030,6035           & 3,4,17 \\ 
    345.504+0.348  & N & N & 1665,1667                     & 3,17 \\
    348.703--1.043 & N & N & 1665                          & 3,17 \\
    NGC6334F       & Y & Y & 1665,1667,1720,6030,6035      & 3,4,5,9 \\
    353.410--0.360 & Y & Y & 1665,1720,4765,6030,6035      & 3,4,5,8,15\\
    9.621+0.196    & Y & N & 1665,1667                     & 3,15 \\
    12.909--0.260  & N & N & 1665,1667,1720                & 3,5,17 \\
    23.010--0.411  & N & N & 1665,1667                     & 6,10 \\
    23.440--0.182  & N & N & 1665,1667                     & 6,10 \\  
    35.201--1.736  & Y & N & 1665,1667,6035                & 4,6,14 \\
    Cep A          &   & N & 1665,1667,4765,6030,6035      & 1,2,16 \\
    NGC7538        & Y & Y & 1665,1667,1720,4765,6035      & 1,2,7,16 \\ \hline
  \end{tabular}
  \label{tab:assoc}
\end{table*}

\subsection{Maser Modelling} \label{sec:modelling}

The methanol maser model of \citet{SD94} can account for maser action
in the 19.9-GHz $2_{1}-3_{0}\ \rm{E}$ transition, as well as in the
other 8 class~II maser transitions for which observations are listed
in Table~\ref{tab:fluxes} \citep{SCG97}.  The same model has been
applied to OH masers in star-forming regions \citep{CSG02}.  When both
molecules are highly abundant in the gas phase, masers of methanol and
OH can be excited under the same model conditions, consistent with the
close associations which are often observed.  Here we examine model
predictions in relation to the newly identified trends for sources
exhibiting 19.9-GHz methanol maser emission.

In the model of Sobolev \& Deguchi, the maser pumping is produced by
infrared emission from warm dust at temperatures exceeding 100~K.  For
methanol this is sufficient to pump the molecules to their second
torsionally excited state; for OH the pumping proceeds via excited
rotational states.  The maser pumping is strongest when the gas is
significantly cooler than the dust, although some masers persist even
when the gas and dust temperatures are comparable.  The masers are
thermally quenched at gas densities exceeding $10^9$~\cc.  Very large
column densities of methanol and OH are required to account for
observed maser brightness temperatures, which can be as great as
$10^{12}$~K.  For plausible maser dimensions, this implies large
abundances relative to hydrogen (e.g. $\Xm=10^{-6}$, $\Xh=5\times
10^{-8}$), which are believed to follow from the evaporation of icy
grain mantles in the neighbourhood of a young stellar object
\citep{HMLD95}.  In the model calculations, the maser intensities can
be enhanced by beaming, and by the amplification of continuum
radiation from an underlying uc\HII\ region.  The principal parameters
governing the behaviour of the masers are the dust temperature \Td\ 
which drives the maser pumping, the gas kinetic temperature \Tk\ and
hydrogen density \nH\ which influence the collisional relaxation, and
the beaming factor \beam\ and specific column density \scd\ of
methanol or OH which determine the optical depth.  In terms of these
parameters the column density of the species in question along the
line of sight is given by $\beam \times \scd \times \DV$, where \DV\ 
is the velocity width of the maser line.

One of the major sources of uncertainty in excitation studies of
methanol has been the influence of collisions, which has until now
been modelled on propensity rules derived from a small number of
experiments \citep{LH74}.  State-to-state rate coefficients for
rotational excitation of methanol have recently been calculated
\citep*{PFD02, PFD04}, and here we report the first methanol maser
calculations to employ the new data.  We include collisional
transitions between the methanol torsional ground state levels,
assuming the collision partners to be 20\% He and 80\% para-H$_2$
(rate coefficients involving ortho-H$_2$ are not yet available).  Data
are provided for methanol energy levels up to J=9, so we extrapolated
these to higher J values using the 1/$\Delta$J propensity rule.  A
full investigation of the effects of the improved collision modelling
on the masers, and of the minor effects of collisions involving higher
torsional states, will be published elsewhere.

The masers are sensitive to the model conditions, switching on and off
as parameters representing physical conditions are varied, with
different combinations of maser lines appearing in different regimes.
The masers at assorted frequencies which are observed in a given
source often appear at the same velocity, and we assume this to mean
that they are simultaneously excited.  In this case, the model can be
used to identify physical conditions which produce the same
combination of masers.  Observations at high spatial resolution are
required to test this assumption by establishing whether or not the
masers coincide in position.

\begin{figure}
  \psfig{file=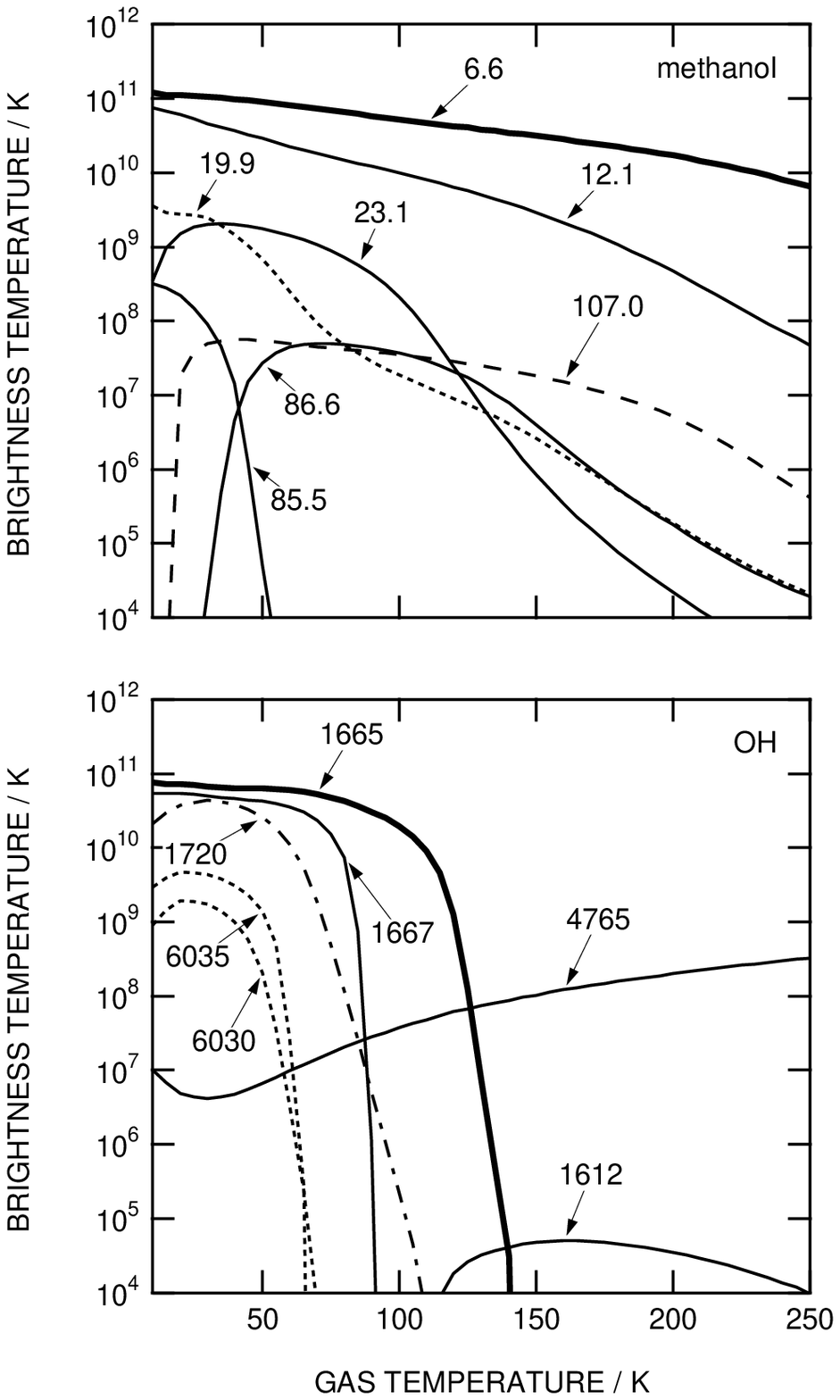,width=8.5cm}
  \caption{Variation of maser brightness temperature \Tb\ with gas kinetic
    temperature \Tk\ for models with dust temperature $\Td=175$~K and
    gas density $\nH=10^7$~\cc.  Top panel shows behaviour of 7
    methanol masers, labelled with the approximate transition
    frequency in GHz, with methanol specific column density
    $\scdm=10^{12}$~\ccs.  Bottom panel shows 7 OH masers, labelled
    with the approximate transition frequency in MHz, with
    $\scdh=10^{11}$~\ccs.}
  \label{fig:model_tk}
\end{figure}

\begin{figure}
  \psfig{file=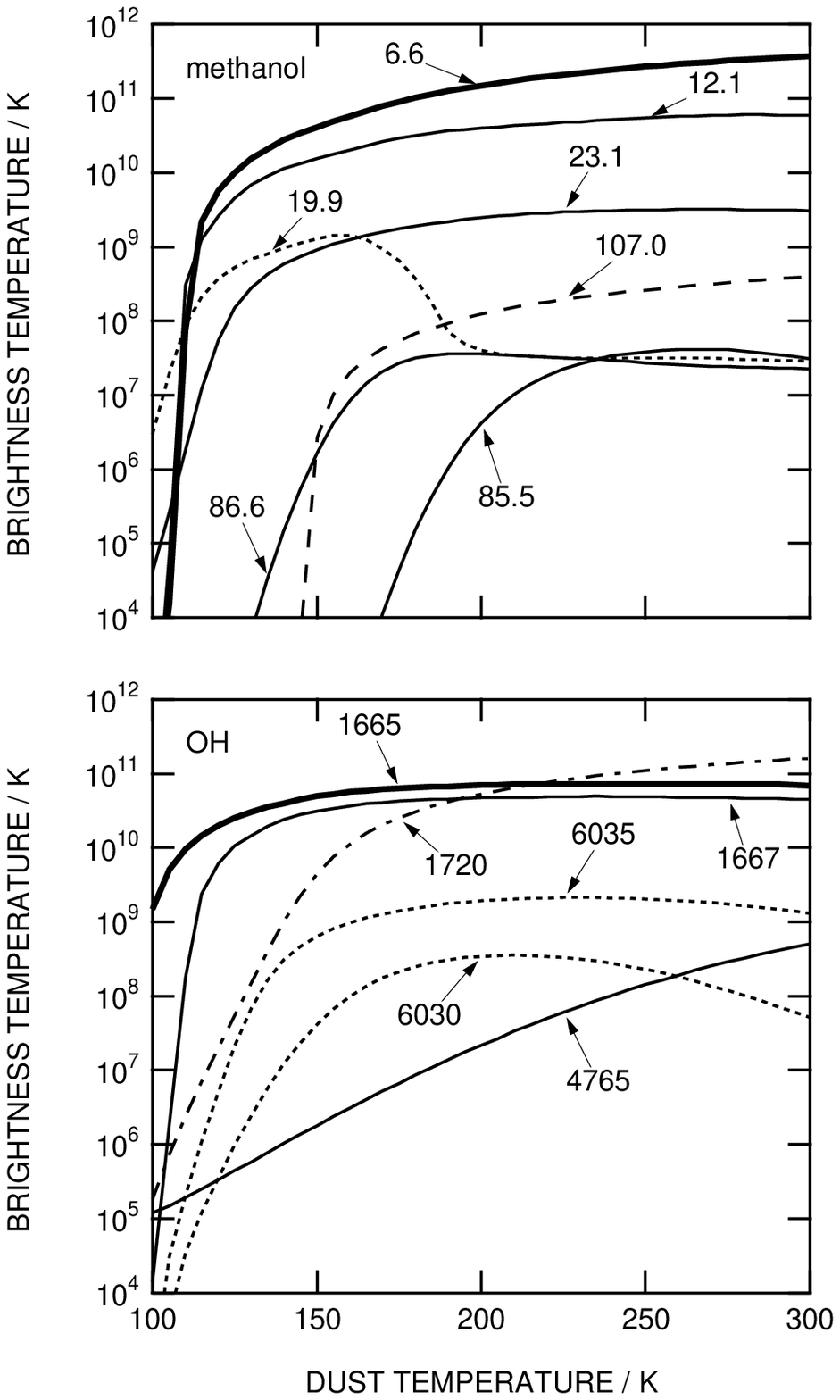,width=8.5cm}
  \caption{Variation of maser brightness temperature \Tb\ with dust
    temperature \Td\ for models with gas kinetic temperature
    $\Tk=50$~K and gas density $\nH=10^7$~\cc.  Top panel shows
    behaviour of 7 methanol masers, labelled with the approximate
    transition frequency in GHz, with methanol specific column density
    $\scdm=10^{12}$~\ccs.  Bottom panel shows 7 OH masers, labelled
    with the approximate transition frequency in MHz, with
    $\scdh=10^{11}$~\ccs.}
  \label{fig:model_td}
\end{figure}

\begin{figure}
  \psfig{file=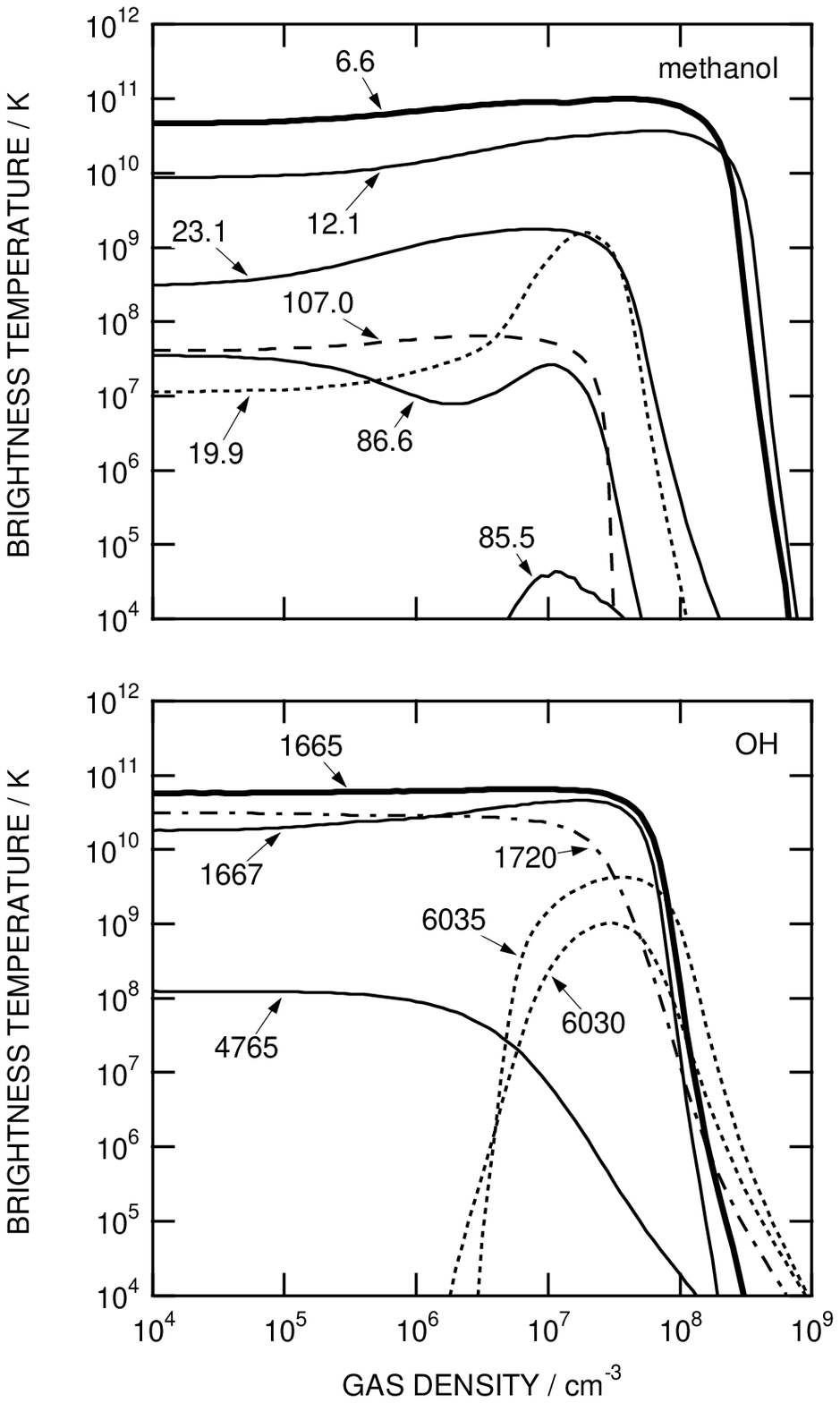,width=8.5cm}
  \caption{Variation of maser brightness temperature \Tb\ with gas density
    \nH\ for models with gas temperature $\Tk=50$~K and dust
    temperature $\Td=175$~K.  Top panel shows behaviour of 7 methanol
    masers, labelled with the approximate transition frequency in GHz,
    with methanol specific column density $\scdm=10^{12}$~\ccs.  Bottom
    panel shows 7 OH masers, labelled with the approximate transition
    frequency in MHz, with $\scdh=10^{11}$~\ccs.}
  \label{fig:model_nh}
\end{figure}

\begin{figure}
  \psfig{file=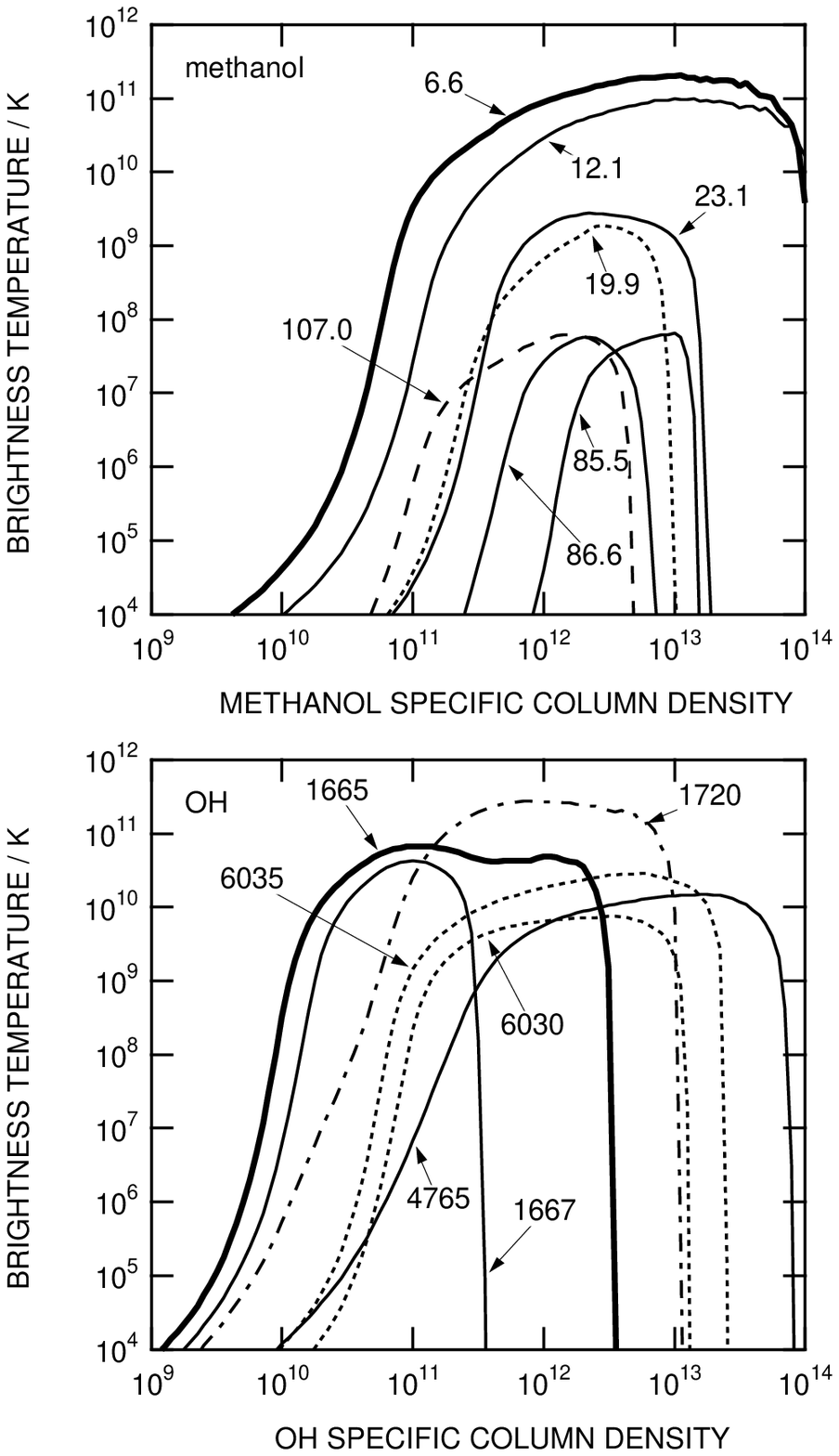,width=8.5cm}
  \caption{Variation of maser brightness temperature \Tb\ with specific column
    density of methanol or OH.  Models have gas kinetic temperature
    $\Tk=50$~K, dust temperature $\Td=175$~K and gas density
    $\nH=10^7$~\cc.  Top panel shows behaviour of 7 methanol masers,
    labelled with the approximate transition frequency in GHz, while
    bottom panel shows 7 OH masers, labelled with the approximate
    transition frequency in MHz.}
  \label{fig:model_scd}
\end{figure}

Figures~\ref{fig:model_tk}-\ref{fig:model_scd} illustrate a range of
model conditions under which methanol and OH masers become active.
Each plot shows the behaviour of maser brightness temperature as a
single model parameter is varied.  The upper panels include data for 7
of the methanol masers for which observations are collected in
Table~\ref{tab:fluxes}: 6.6, 12.1, 19.9, 23.1, 85.5, 86.6 and
107.0~GHz.  Curves for 108.8 and 156.6~GHz are omitted in the
interests of clarity -- these resemble the curves for 107.0~GHz, but
are generally weaker.  The lower panels show the behaviour of the 7
most prevalent OH masers at 1612, 1665, 1667, 1720, 4765, 6030 and
6035~MHz.  The strongest and most widespread observed masers (6.6-GHz
methanol and 1665-MHz OH) are shown as bold traces; in most cases
these are also the strongest masers in the models.  The calculations
were based around conditions which illustrate significant methanol
maser action at 19.9~GHz: gas temperature $\Tk=50$~K, dust temperature
$\Td=175$~K, hydrogen density $\nH=10^7$~\cc, and methanol specific
column density $\scdm=10^{12}$~\ccs.  The OH specific column density
is $\scdh=10^{11}$~\ccs.  In these calculations the beaming factor was
$\beam=10$.  Since 19.9-GHz emission is detected in sources where the
methanol masers are projected against radio continuum emission, we
also included background continuum radiation in the modelling.  The
\HII\ continuum spectrum has temperature $\Te [1-\rm{exp}(-(\fe /
\nu)^2)]$~K, where $\Te=10^{4}$~K and $\fe=12$~GHz, and is
geometrically diluted by a factor $\WHII=0.002$.  Apart from the
choice of parameter values and the new collision rate coefficients 
for methanol, the model calculations are identical to
those described in full detail in Cragg \etal\ (2002).

For all but one of the sources in Table~\ref{tab:fluxes} the flux
density at 107.0~GHz exceeds that at 19.9~GHz, the exception being the
very strong 19.9-GHz maser in NGC~6334F.  To compare these flux
densities $S$ with the model calculations of maser brightness
temperature \Tb\ requires an estimate of the maser source angular
extent $\Omega$, since $S$ depends on $\nu^2 \Omega \Tb$.  In the
absence of this information, we seek models which can reproduce the
observed flux density ratios, assuming that the masers of different
frequency in the same source have a common size.  When the frequency
factor is taken into account, the observations suggest that for the
sources where the 19.9-GHz emission is detected (excepting NGC6334F),
the ratio of maser brightness temperatures for the component at the
19.9-GHz peak velocity lies in the range $0.07 \leq
\Tb(19.9)/\Tb(107.0) \leq 17$.  Thus the models which best represent
the sources in our sample, all of which have detectable 107.0-GHz
maser emission, will be those in which the 107.0-GHz maser is
relatively strong.  Conditions representing the sources with
associated 19.9-GHz emission are likely to be in the vicinity of the
intersection of the 19.9-GHz (dotted) and 107.0-GHz (dashed)
traces.  The correlation between 19.9-GHz methanol emission and 6030-
and 6035-MHz OH emission in the model can be examined by comparing the
dotted traces in the upper and lower panels of
Figs~\ref{fig:model_tk}-\ref{fig:model_scd}.

Fig.~\ref{fig:model_tk} shows the variation with gas temperature \Tk.
The strongest methanol masers are, as observed, at 6.6 and 12.1~GHz,
with the 19.9-GHz maser being most favoured at gas temperatures below
50~K.  As the 23.1-GHz maser is also rather strong at gas temperatures
up to 100~K, its nondetection (Cragg \etal\ 2004) can set useful
limits on model parameters, but note that the 23.1-GHz survey was made
with lower sensitivity than obtained in the current investigation at
19.9~GHz.  Based on this plot, our observed flux density ratios would
suggest that NCG~6334F may have rather low gas temperature \citep[$<
50$~K, consistent with the earlier modelling of][]{CSECGSD01} , the
other 19.9-GHz maser sources may have temperatures in the range
$50-150$~K, while the sources in which no 19.9-GHz maser was detected
may have temperatures in the same range or higher.  Gas temperatures
above 100~K in the majority of sources would be consistent with the
scarcity of masers at 23.1, 85.5, 86.6, 108.8 and 156.6~GHz.  Note
that such one-dimensional analysis is very simplistic, since different
choices of gas density, dust temperature, methanol abundance etc. will
influence the conclusions drawn, but this nevertheless gives a useful
starting point for future multidimensional fitting of the model to the
observations.  The lower panel of Fig.~\ref{fig:model_tk} shows that
the 6030-, 6035- and 1720-MHz OH masers are also favoured by gas
temperatures below 75~K, in keeping with their observed correlation
with 19.9-GHz methanol masers.

Fig.~\ref{fig:model_td} shows the variation with dust temperature \Td.
The 19.9-GHz maser is strongest for dust temperatures $100-200$~K.
This is consistent with the infrared observations of
\citet{DPT00,DWPPT02} which suggest dust temperatures in the range
$120-160$~K for our 19.9-GHz methanol maser sources, as noted in
section~\ref{sec:indiv}.  The various transitions have slightly
different threshold temperatures for maser activity.  For example the
107.0-GHz methanol maser requires dust temperature above 150~K to turn
on under the chosen conditions, and is brighter than the 19.9-GHz
maser at temperatures above 200~K.  The 6030-, 6035- and 1720-MHz OH
masers are also strong at dust temperatures above 150~K.

Fig.~\ref{fig:model_nh} shows the variation with gas density \nH.  All
the masers are thermally quenched at the highest density $10^9$~\cc.
Under the conditions illustrated the 107.0-GHz maser is quenched at
$\nH>10^{7.5}$~\cc, while the 19.9-GHz line is strongest for densities
$10^{6.5} - 10^8$~\cc.  These conditions also favour maser activity in
the 6030-, 6035- and 1720-MHz OH transitions.

Fig.~\ref{fig:model_scd} shows the variation with specific column
density \scd\ of methanol and OH.  The calculations on which
Figs~\ref{fig:model_tk}-\ref{fig:model_nh} are based have
$\scdm=10\times\scdh$, following Cragg \etal\ (2002) where it was
established that this factor accounts well for the observed intensity
ratios of the dominant 6.6-GHz methanol and 1665-MHz OH masers.
Fig.~\ref{fig:model_scd} shows that this factor also provides
substantial common ground between the 19.9-GHz methanol and 6030-,
6035- and 1720-MHz OH masers, in keeping with the correlation
identified by our observations.  The plotted brightness temperatures
suggest that our sample of 107.0-GHz methanol maser sources have
\scdm\ in the range $10^{11} - 10^{12.5}$~\ccs.

We may conclude that the maser models are able to account
qualitatively for the detection of 19.9-GHz methanol masers in 9 out
of 25 of the 107.0-GHz sources, and are also able to explain the
correlation found between methanol emission at 19.9~GHz and OH masers
at 6035~MHz.  The model calculations suggest that the 19.9-GHz maser
sources have gas temperatures around 50~K, dust temperatures
150-200~K, gas density $10^{6.5}-10^{7.5}$~\cc and methanol specific
column density $10^{11}-10^{12.5}$~\ccs, while the sources with no
19.9-GHz maser detected may be higher in gas or dust temperature or
lower in gas density.  As these estimates were determined in a
simplistic fashion by varying only one model parameter at a time, they
cannot be considered definitive.  A more quantitative comparison
between model calculations and observations in individual sources can
further constrain the physical conditions which give rise to the
observed masers.  This is best undertaken after observations at high
spatial resolution have clarified the associations between the
different masers.  The exceptional sources W3(OH), 345.010+1.792 and
NGC6334F have been the subject of previous detailed modelling of this
type \citep{SSECMOG01,CSECGSD01}, and fitting of the maser model to
observations of the 25 107.0-GHz maser sources is the subject of
ongoing work.  For this purpose further model calculations are
required to explore the multidimensional parameter space in the
regions identified above.

\section{Conclusions}

We have made the first sensitive search for the 19.9-GHz transition of
methanol in the southern hemisphere and detected 6 new sources, taking
the total number to 9.  There is a strong correlation between sources
which show 19.9-GHz methanol emission and those which both have the
class~II methanol masers projected against radio continuum and have
associated 6035-MHz OH.  This suggests that the 19.9-GHz methanol
masers trace a narrow range of physical conditions, probably
associated with a specific evolutionary phase.  We have made a
preliminary examination of this association in the model of
\citet{CSG02} and find it to be consistent with a gas temperatures of
50~K, dust temperatures of 150-200~K and gas densities of
$10^{6.5}-10^{7.5}$ \cc, although further modelling is required to
more thoroughly sample the parameter space.

\section*{Acknowledgements}

This research has made use of NASA's Astrophysics Data System Abstract
Service.  This research has made use of the SIMBAD database operated
at CDS, Strasbourg, France.  The authors thank David Flower for making
methanol rate coefficient data available prior to publication.
Financial support for this work was provided by the Australian
Research Council.  AMS was supported by the Ministry of Industry,
Science and Technology of the Russian Federation (Contract No.
40.022.1.1.1102).

\end{document}